\documentclass[twocolumn,  tighten,  twocolappendix]{aastex631} 

\shortauthors{Yoon et al. (2024)}

\begin{document}

\title{The Most Massive Early-type Galaxies Exhibit Tidal Features More Frequently in Lower-density Environments}

\email{yyoon@kasi.re.kr}

\author[0000-0003-0134-8968]{Yongmin Yoon}
\affiliation{Korea Astronomy and Space Science Institute (KASI), 776 Daedeokdae-ro, Yuseong-gu, Daejeon 34055, Republic of Korea}

\author[0000-0002-1710-4442]{Jae-Woo Kim}
\affiliation{Korea Astronomy and Space Science Institute (KASI), 776 Daedeokdae-ro, Yuseong-gu, Daejeon 34055, Republic of Korea}

\author[0000-0002-9434-5936]{Jongwan Ko}
\affiliation{Korea Astronomy and Space Science Institute (KASI), 776 Daedeokdae-ro, Yuseong-gu, Daejeon 34055, Republic of Korea}
\affiliation{University of Science and Technology, Gajeong-ro, Daejeon 34113, Republic of Korea}

\begin{abstract}
The most massive early-type galaxies (ETGs) are known to form through numerous galaxy mergers. Thus, it is intriguing to study whether their formation in low-density environments, where nearby companions are almost absent, is associated with mergers, which are directly traced by tidal features. Using the 436 most massive ETGs with $M_\mathrm{star}>10^{11.2}\,M_{\odot}$ at $z<0.04$, we determine the variation in the fraction of massive ETGs with tidal features ($f_T$) across different environments and verify whether the most massive ETGs commonly have tidal features in very low density environments. Our main discovery is that the most massive ETGs exhibit tidal features more frequently in lower-density environments. In the highest-density environments, like galaxy clusters, $f_T$ is $0.21\pm0.06$, while in the lowest-density environments it triples to $0.62\pm0.06$. This trend is stronger for more extremely massive ETGs, with $f_T$ reaching $0.92\pm0.08$ in the lowest-density environments. One explanation for our finding is that the most massive ETGs in lower-density environments have genuinely experienced recent mergers more frequently than their counterparts in higher-density environments, suggesting that they possess extended formation histories that continue into the present. Another possibility is that tidal features last shorter in denser environments owing to external factors inherent in these environments. Our additional findings that massive ETGs with bluer $u-r$ colors are a more dominant driver of our main discovery and that dust lanes are more commonly observed in massive ETGs in low-density environments imply that gas-abundant mergers primarily contribute to the increased rate of recent mergers in low-density environments.
\end{abstract}
\keywords{Early-type galaxies (429); Galaxy environments (2029); Galaxy mergers (608); Giant galaxies (652); Tidal tails (1701)}

\section{Introduction}\label{sec:intro}

Early-type galaxies (ETGs) are preferentially detected in high-density environments, such as galaxy clusters, whereas late-type galaxies show an inverse trend, being more common in low-density environments \citep{Dressler1980,Postman1984,Goto2003,Houghton2015,Pfeffer2023}. This morphology--density relation indicates that galaxy evolution is associated with the environments in which galaxies reside. 

Although ETGs are less common in low-density environments, they are still present in isolated environments, where there are virtually no nearby companion galaxies \citep{Reduzzi1996,Mulchaey1999,Colbert2001,Reda2004,Reda2005,Reda2007,Stocke2004,Hau2006,Hernandez-Toledo2008,Lacerna2016,Rampazzo2020}. Several studies have examined the properties of ETGs in such isolated environments and compared them to those in high-density environments. Many of them found that ETGs in isolated environments share similar properties with those in dense environments, such as color, sizes, star formation rates, kinematics, luminosity function, and scaling relations between the properties \citep{Mulchaey1999,Reda2004,Reda2005,Reda2007,Stocke2004,Hau2006,Lacerna2016}. This implies that the crucial formation mechanisms for ETGs are likely common across various environments, although the cumulative number of galaxies influenced by these mechanisms over the history of the universe should be higher in denser environments, contributing to the observed morphology--density relation. 

Galaxy mergers are likely to be a universally important mechanism for the formation and evolution of ETGs across various environments, as they can establish the fundamental properties of ETGs \citep{Baugh1996,Christlein2004,DeLucia2006,DeLucia2007,Wilman2013}. For instance, galaxy mergers can lead to the creation of red and quiescent remnants that no longer actively form new stars \citep{Springel2005,Hopkins2008,Brennan2015}, as the available cold gas is rapidly depleted through intense star formation events during the mergers \citep{Hernquist1989,Mihos1996,Springel2005}.

Under the current understanding of the universe, the galaxy merger rate is known to depend on galaxy environments and cosmological epochs \citep{Kauffmann1996,Gottlober2001,DeLucia2006,Niemi2010,Yoon2017,Yoon2023b}, which is also supported by observational evidence \citep{vanDokkum1999,Kuntschner2002,Reda2004,Reda2005,Reda2007,Collobert2006}. Specifically, low-redshift ETGs in cluster-like high-density environments are more likely to have experienced mergers and consequent evolution at early epochs of $z\gtrsim2$, when galaxies within the halos had velocities low enough to allow frequent mergers. However, in the low-redshift universe, galaxy mergers are inhibited by the high relative velocities within fully developed, massive halos with a high density of galaxies. By contrast, ETGs in low-density regions, including those in isolated environments, tend to have been assembled more recently at $z\lesssim1$. In other words, they have prolonged formation histories extending to the current epoch, undergoing continual, though less frequent, mergers in low-density environments.

If so, it is expected that in the low-redshift universe tidal features are frequently detected around ETGs in low-density environments, or even more so than around ETGs in dense environments, since tidal features are remnants of stellar debris created by galaxy mergers \citep{Quinn1984,Barnes1988,Hernquist1992,Feldmann2008}, serving as direct observational evidence of recently occurred mergers. Indeed, several observational studies have revealed that traces of recent mergers, such as tidal features, are more commonly found around ETGs in lower-density or isolated environments \citep{Reduzzi1996,Colbert2001,Michard2004,Reda2005,Tal2009}.

Galaxy mergers are more essential for the formation of the most massive ETGs with stellar masses exceeding $M_\mathrm{star}>10^{11}\,M_{\odot}$ than for the formation of less massive ones. The assembly of mass in these massive ETGs is largely contributed by ex-situ sources \citep{Dubois2016,Davison2020}, typically through multiple mergers, including the accretion of small galaxies \citep{Oogi2013,Rodriguez-Gomez2015,Yoon2017,Husko2022}. Considering the significance of multiple mergers in forming the most massive ETGs, it is intriguing to investigate whether the assembly of such massive ETGs in low-density environments, where nearby companion galaxies are almost absent, is also related to galaxy mergers.

In this study, we verify for the first time whether the most massive ETGs with $\log(M_\mathrm{star}/M_{\odot})>11.2$ frequently exhibit tidal features in very low density environments. We also qualitatively determine the variation in the fraction of massive ETGs with tidal features across different environments, using a large sample of the 436 most massive ETGs at $z<0.04$, for which deep images are available from the Dark Energy Spectroscopic Instrument (DESI) Legacy Imaging Survey \citep{Dey2019}. By doing so, we enhance our understanding of the dependence of merger rates on environments in the low-redshift universe and also discuss the persistence of tidal features across different environments.

In this study, the cosmological parameters are set as follows: $H_0=70$ km s$^{-1}$ Mpc$^{-1}$ for the Hubble constant, $\Omega_{\Lambda}=0.7$ for the dark energy density, and $\Omega_m=0.3$ for the matter density.
\\

\begin{figure*}
\includegraphics[width=\linewidth]{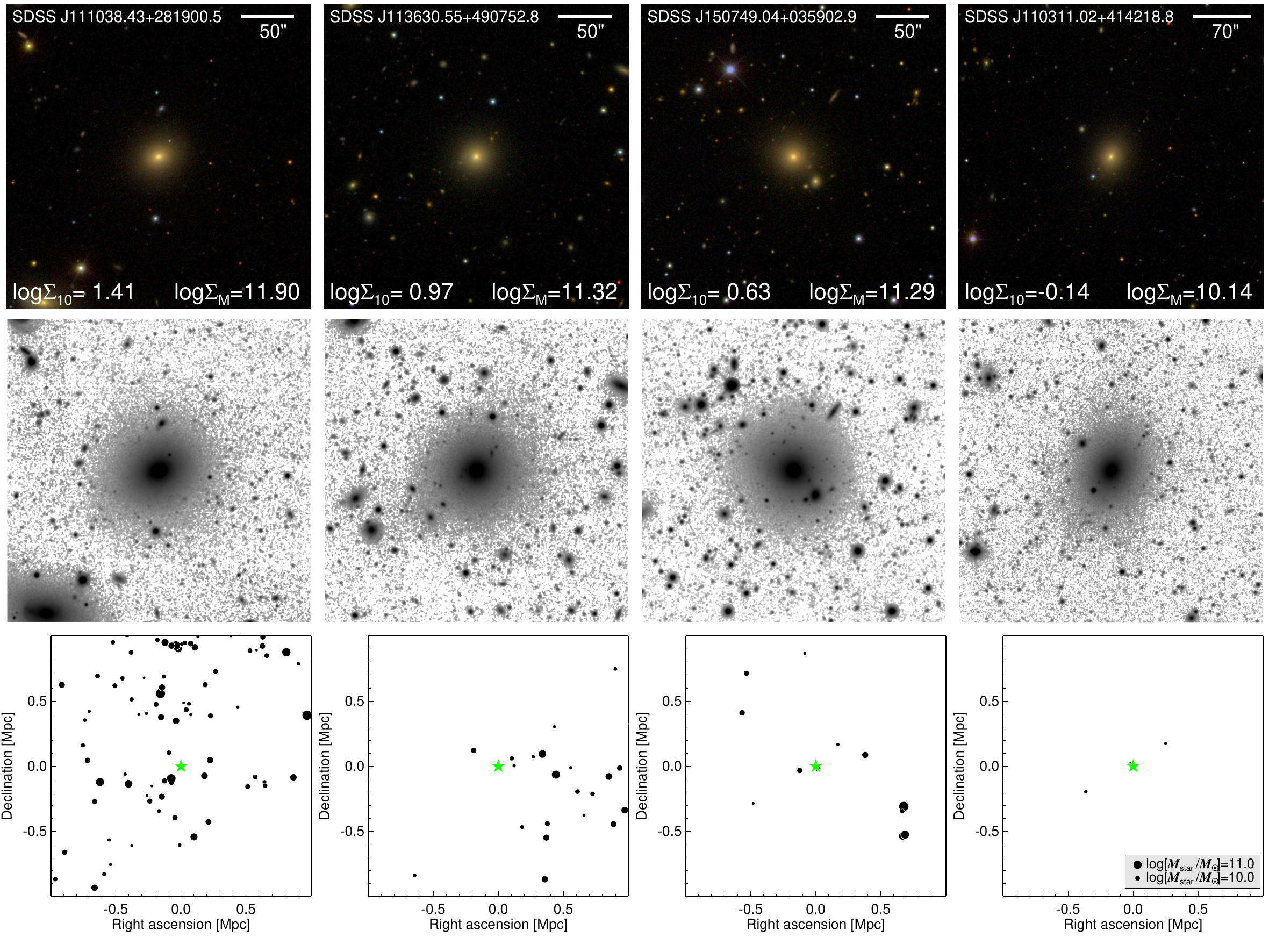}
\centering
\caption{Examples of massive ETGs without tidal features. First row: color images from SDSS. The galaxy ID, $\log\Sigma_{10}$, and $\log\Sigma_{M}$ are displayed in the color images. The units of $\Sigma_{10}$ and $\Sigma_{M}$ are Mpc$^{-2}$ and $M_{\odot}\,$Mpc$^{-2}$, respectively. The horizontal bar in the color image denotes the angular scale of the image. Second row: $r$-band deep images of the DESI Legacy Survey. The angular scale of the deep image is identical to that of the color image in the first row. Third row: two-dimensional maps illustrating the spatial distribution of galaxies within $\sim1$ Mpc from each massive ETG. The green star at the center of the map represents the location of the massive ETG, whereas the black filled circles denote other galaxies in the environment. The size of each circle indicates the galaxy's stellar mass, as shown in the legend in the bottom right panel. The ETGs are arranged such that the galaxy density decreases from the left to the right panels.
\label{fig:ex_n}}
\end{figure*} 

\begin{figure*}
\includegraphics[width=\linewidth]{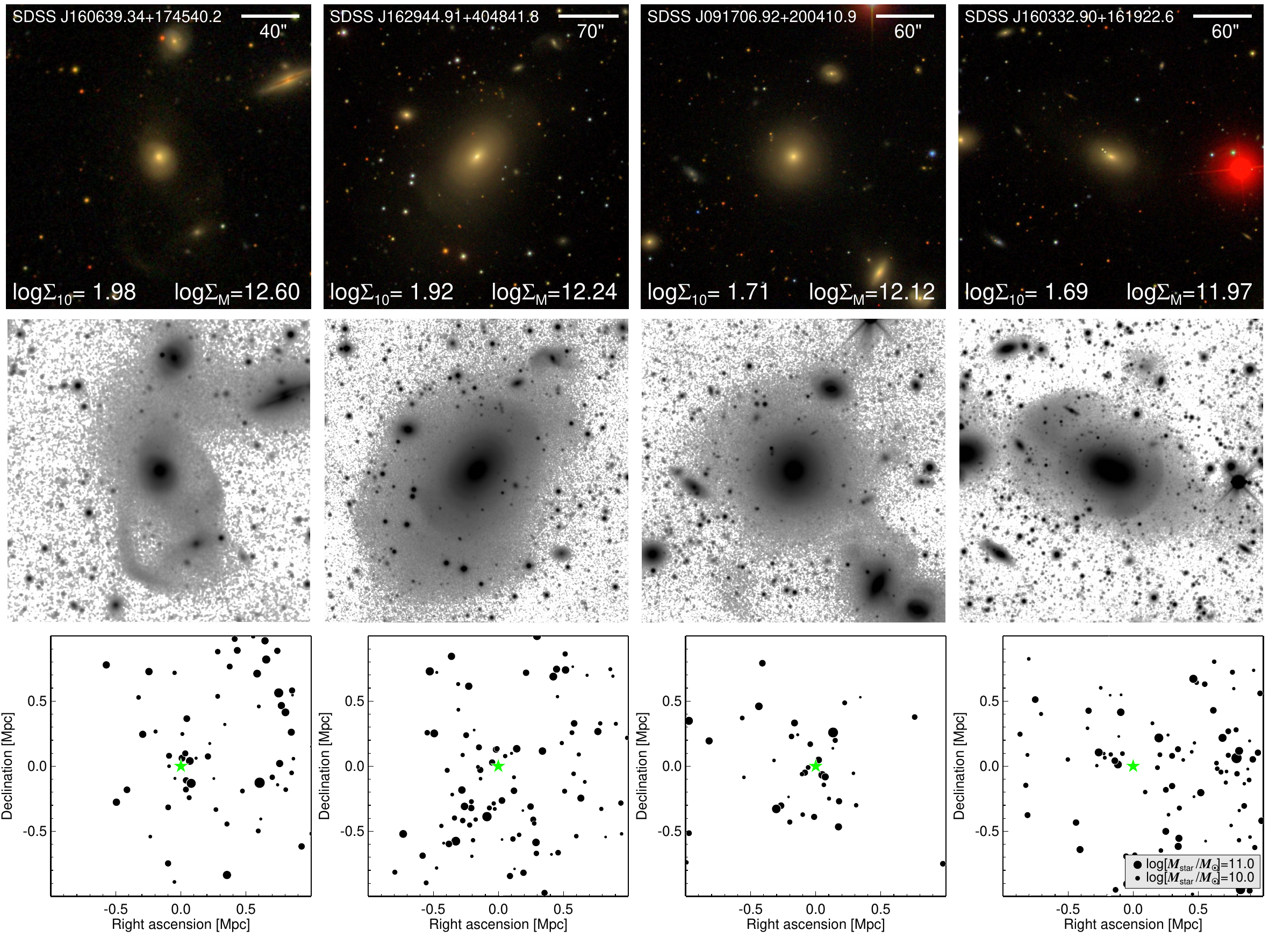}
\centering
\caption{Examples of massive ETGs with tidal features, which are located in the environments with $\Sigma_{10}>48$ (or $\log\Sigma_{M}>11.9$), where the units of $\Sigma_{10}$ and $\Sigma_{M}$ are Mpc$^{-2}$ and $M_{\odot}\,$Mpc$^{-2}$, respectively. The other descriptions for this figure are the same as in Figure \ref{fig:ex_n}.
\label{fig:ex_1}}
\end{figure*} 

\begin{figure*}
\includegraphics[width=\linewidth]{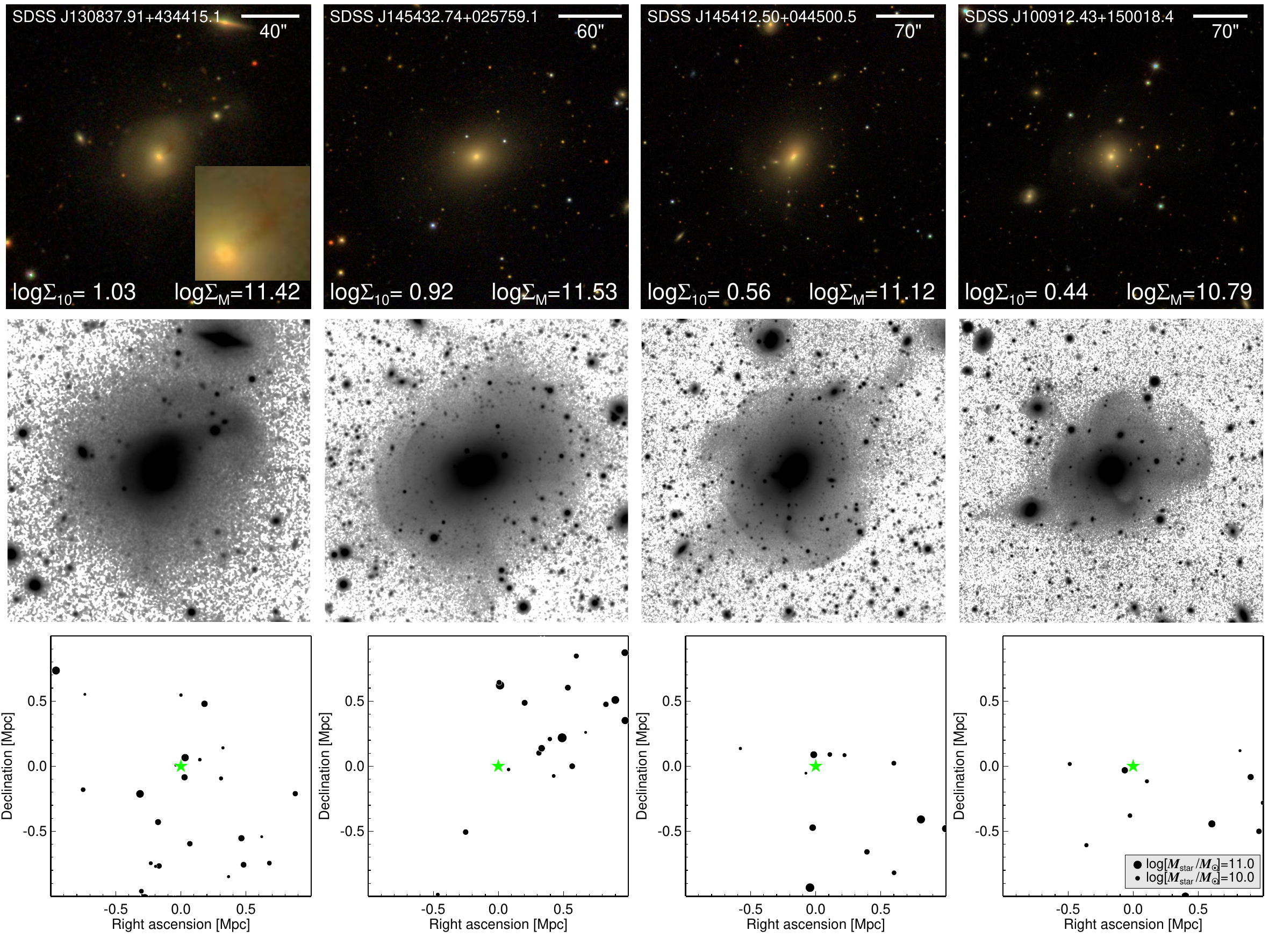}
\centering
\caption{Examples of massive ETGs with tidal features, which are located in the environments with $2.7<\Sigma_{10}<11$ (or $10.7<\log\Sigma_{M}<11.6$), where the units of $\Sigma_{10}$ and $\Sigma_{M}$ are Mpc$^{-2}$ and $M_{\odot}\,$Mpc$^{-2}$, respectively. The inset in the first color image magnifies the main body of the galaxy to clearly show the presence of a dust lane. The other descriptions for this figure are the same as in Figure \ref{fig:ex_n}.
\label{fig:ex_2}}
\end{figure*} 

\begin{figure*}
\includegraphics[width=\linewidth]{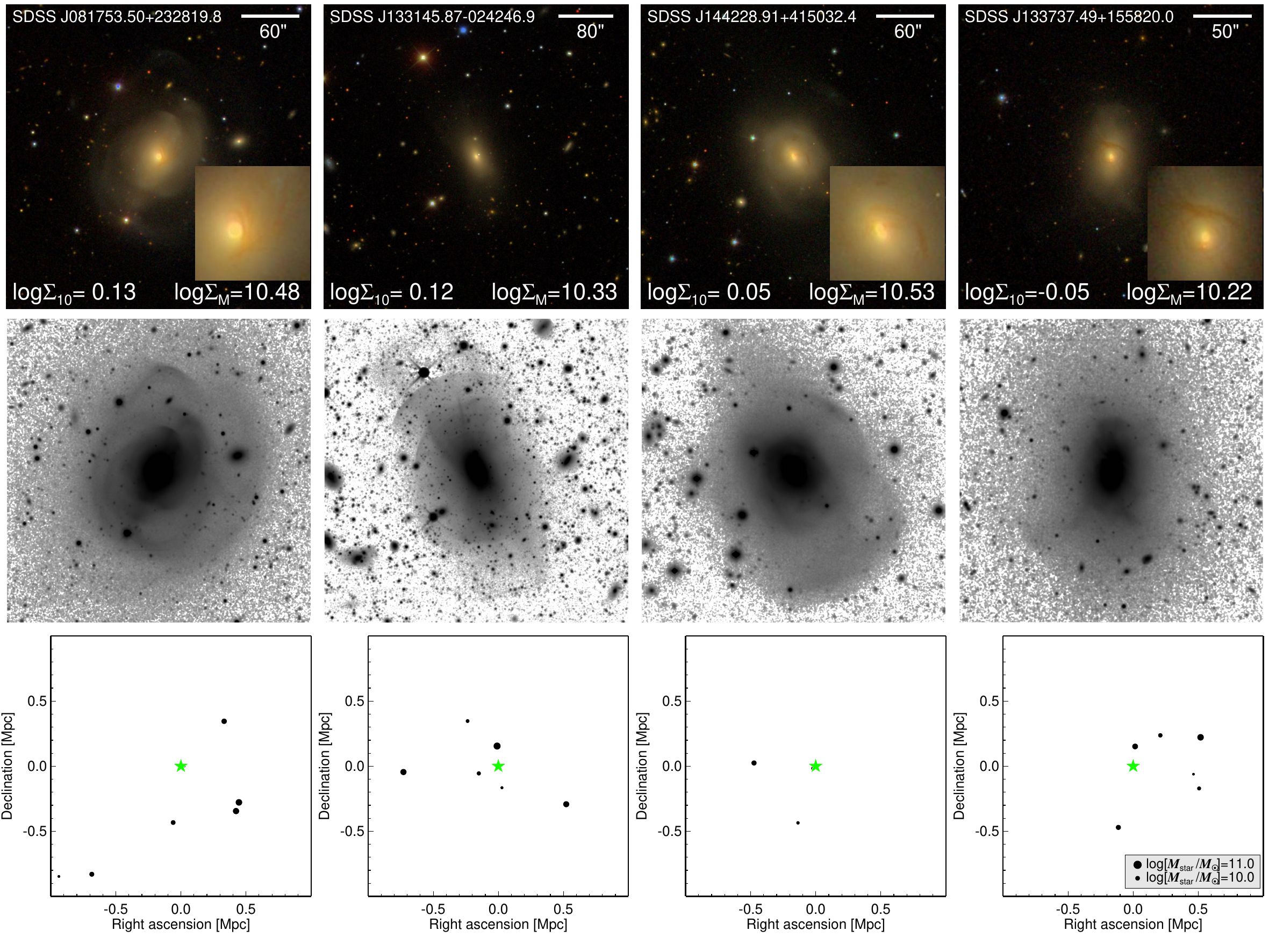}
\centering
\caption{Examples of massive ETGs with tidal features, which are located in the environments with $0.8<\Sigma_{10}<1.4$ (or $10.2<\log\Sigma_{M}<10.6$), where the units of $\Sigma_{10}$ and $\Sigma_{M}$ are Mpc$^{-2}$ and $M_{\odot}\,$Mpc$^{-2}$, respectively. The inset images in the first row magnify the main body of each galaxy to clearly show the presence of dust lanes. The other descriptions for this figure are the same as in Figure \ref{fig:ex_n}.
\label{fig:ex_3}}
\end{figure*} 

\begin{figure*}
\includegraphics[width=\linewidth]{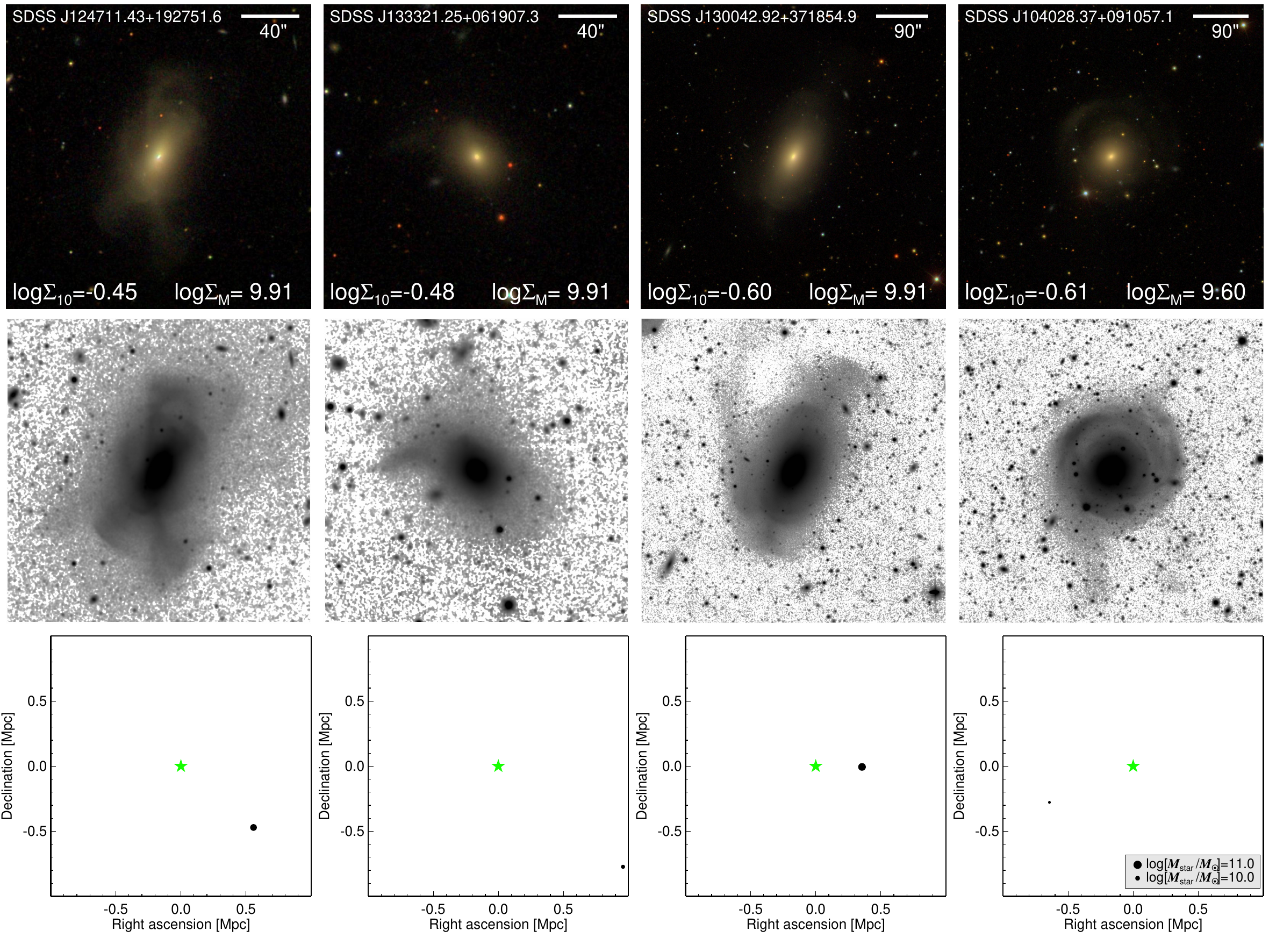}
\centering
\caption{Examples of massive ETGs with tidal features, which are located in the environments with $\Sigma_{10}<0.4$ (or $\log\Sigma_{M}<10.0$), where the units of $\Sigma_{10}$ and $\Sigma_{M}$ are Mpc$^{-2}$ and $M_{\odot}\,$Mpc$^{-2}$, respectively. The other descriptions for this figure are the same as in Figure \ref{fig:ex_n}.
\label{fig:ex_4}}
\end{figure*}

\section{Sample}\label{sec:sample}
The sample of the most massive ETGs and galaxies used for defining environments is sourced from the NASA-Sloan Atlas (NSA) catalog. This catalog provides almost all documented redshifts up to $z=0.15$ for galaxies covered by the Sloan Digital Sky Survey (SDSS) Data Release 11, containing about 640,000 galaxies.\footnote{For information on the image analysis conducted for this catalog, from which galaxy properties are derived, see \citet{Blanton2011}.}

We constrain the parent sample for massive ETGs to a redshift range of $0.01<z<0.04$. This range enables us to compute surface number and stellar mass densities of galaxies, using a galaxy sample whose stellar masses are nearly complete down to $\log(M_\mathrm{star}/M_{\odot})\approx9.4$ (Section \ref{sec:env}). Specifically, the $r$-band magnitude threshold for the spectroscopic target selection (the main galaxy sample; \citealt{Strauss2002}), set at $m_r\approx17.77$, is equivalent to a stellar mass of $\log(M_\mathrm{star}/M_{\odot})\sim9.15$ at the upper limit of $z\approx0.04$. In addition, at $z\sim0.04$, about $80\%$ of the galaxies with $m_r\approx17.77$ have stellar masses less than $\log(M_\mathrm{star}/M_{\odot})=9.4$, and over $\sim90\%$ possess stellar masses below $\log(M_\mathrm{star}/M_{\odot})=9.5$. At the median redshift ($z\sim0.03$) of our most massive ETG sample, $m_r\approx17.77$ is equivalent to $\log(M_\mathrm{star}/M_{\odot})\sim9.0$, and more than $\sim95\%$ of the galaxies with $m_r\approx17.77$ have stellar masses below $\log(M_\mathrm{star}/M_{\odot})=9.4$. The number of galaxies with $\log(M_\mathrm{star}/M_{\odot})\ge9.4$ in $0.01<z<0.04$ is 53,641 in the NSA catalog.

To guarantee reliable measurements of galaxy environments, the parent sample for massive ETGs is selected from the 53,641 galaxies, including only those that are located in regions (defined as the area within a $0.5\degr$ radius around each galaxy\footnote{The radius $0.5\degr$ corresponds to $\sim1.1$ Mpc at the median redshift ($z\sim0.03$) of our massive ETG sample.}) where the spectroscopic completeness exceeds $70\%$ for galaxies with $m_r<17.77$, and that are positioned at least $\sim0.3\degr$ away from the survey boundary. The parent sample for massive ETGs then comprises 40,515 galaxies. The selection criterion based on spectroscopic completeness predominantly includes galaxies in the main galaxy sample survey area \citep{Strauss2002}, which exhibit a typical spectroscopic completeness of $\sim90\%$ for galaxies with $m_r\lesssim17.77$. Indeed, the spectroscopic completeness around the massive ETGs used here is on average $89\%$, with $94\%$ of these massive ETGs located in regions where the spectroscopic completeness exceeds $80\%$.

Throughout this study, we use stellar masses from the NSA catalog, which are derived from two-dimensional S{\'e}rsic fits. This catalog also provides stellar masses based on elliptical Petrosian fluxes. The two stellar mass estimates derived by the two different methods correlate almost perfectly with each other, though the masses from the S{\'e}rsic fits are on average more massive by 0.1 dex. The Spearman's correlation coefficient between the two estimates for 40,515 galaxies in the parent sample is 0.96, and the scatter around the linear relation between them\footnote{$\log(M_{\mathrm{star},e}/M_{\odot})=0.68+0.93\log(M_{\mathrm{star},s}/M_{\odot})$, where $M_{\mathrm{star},e}$ and $M_{\mathrm{star},s}$ denote the stellar masses based on elliptical Petrosian fluxes and two-dimensional S{\'e}rsic fits, respectively.} is 0.07. Among the galaxies in the parent sample, $2.7\%$ deviate from the relation by more than five times the scatter. We do not take into account those galaxies with large deviations when defining the sample of massive galaxies to ensure the selection of genuinely massive ones without spurious mass estimations.

Galaxies with $\log(M_\mathrm{star}/M_{\odot})>11.2$ are selected for inclusion in the sample of massive galaxies. Such massive galaxies, particularly massive ETGs, are known to have been greatly influenced by mergers in their formation histories, resulting in their warped/bent scaling relations \citep{Hyde2009,Bernardi2011a,Bernardi2011b,Yoon2017,YP2022}. The number of these massive galaxies is 659. By visually inspecting the morphologies of the galaxies through color images of SDSS and the DESI Legacy Survey, we identify 451 ETGs among the 659 massive galaxies. Finally, 15 ETGs are excluded because their images are of poor quality, a result of being too close to bright sources (Section \ref{sec:tidal}). As a result, the massive ETG sample of this study comprises 436 ETGs with $\log(M_\mathrm{star}/M_{\odot})>11.2$. The first rows of Figures \ref{fig:ex_n}--\ref{fig:ex_4} display the color images of several massive ETGs. 
\\

\begin{figure}
\includegraphics[width=\linewidth]{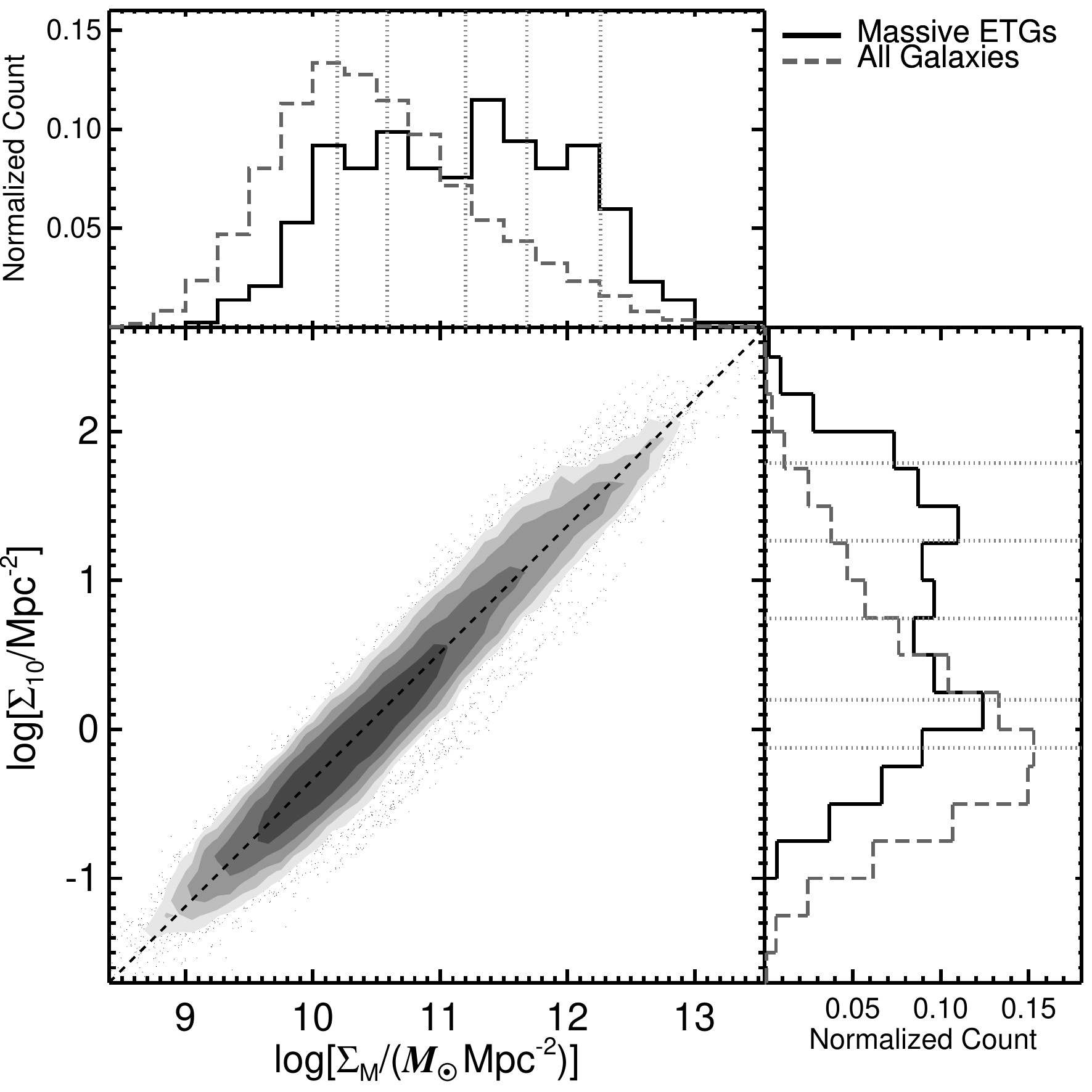}
\caption{Comparison between the logarithmic values of $\Sigma_{10}$ and $\Sigma_{M}$ for 40,515 galaxies with $\log(M_\mathrm{star}/M_{\odot})\ge9.4$ within $0.01<z<0.04$. The filled gray contours represent the density of data points, with darker contours indicating higher densities. The black dashed line crossing the contours denotes a line fitted to the distribution of data points. The line is described by the equation $\log\Sigma_{10} = 0.85\log\Sigma_{M}-8.85$. Here, only data points within $5\sigma$ from the line are displayed, with $\sigma$ being 0.17 dex along the $y$-axis. Also shown are histograms of the distributions of $\log\Sigma_{10}$ and $\log\Sigma_{M}$ for both massive ETGs with $\log(M_\mathrm{star}/M_{\odot})>11.2$ and the 40,515 galaxies. The histograms are normalized so that the total count equals unity. In the histogram panels, the dotted lines dividing $\Sigma_{10}$ or $\Sigma_{M}$ indicate the six environment bins used in Section \ref{sec:results}.
\label{fig:den}}
\end{figure} 

\begin{figure}
\includegraphics[width=\linewidth]{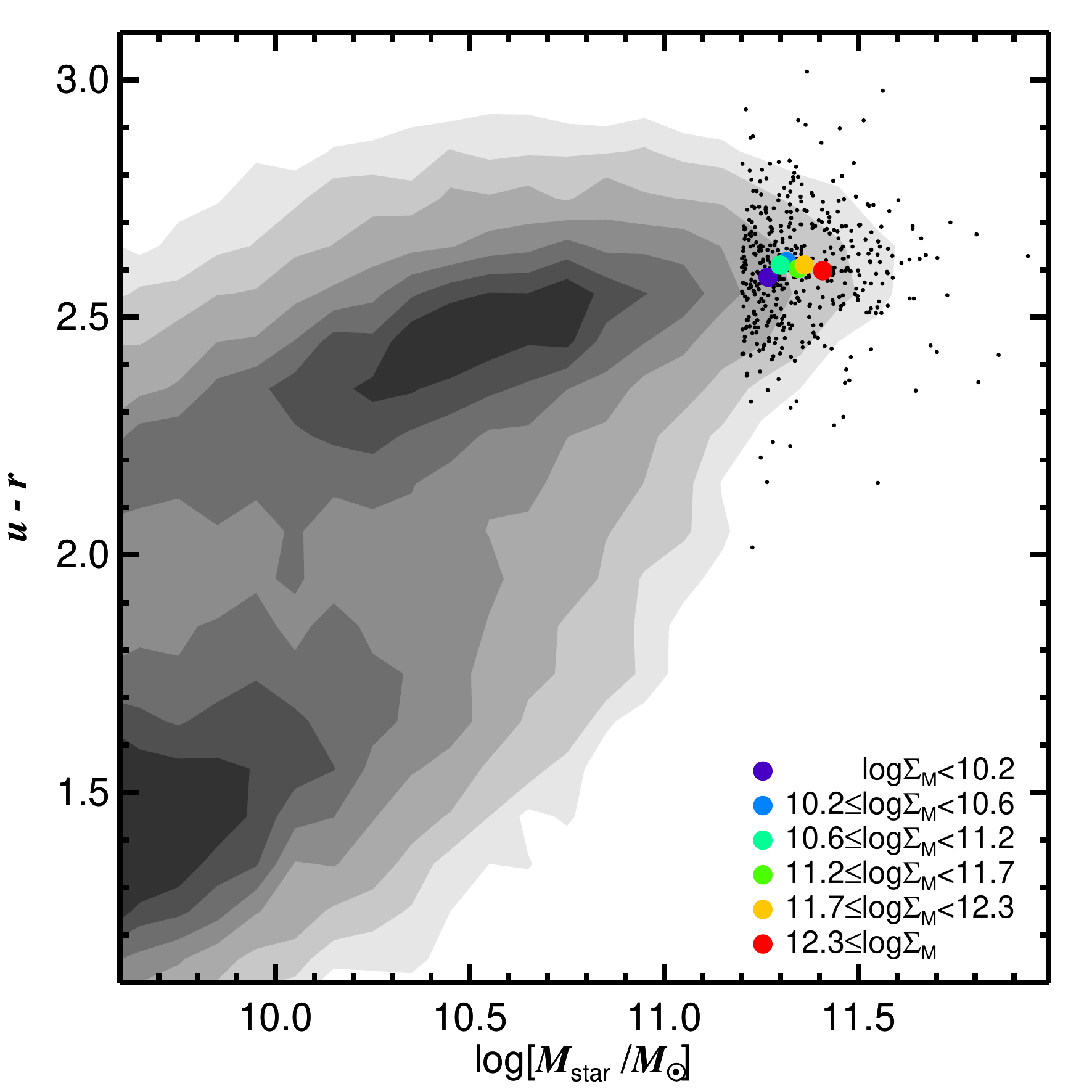}
\caption{Diagram of $u-r$ color vs. stellar mass. The contours denote the density of all galaxies within $0.01<z<0.04$, with darker contours indicating higher densities. Massive ETGs with $\log(M_\mathrm{star}/M_{\odot})>11.2$ are plotted as black circles. We represent the median values of colors and stellar masses for massive ETGs in six different environments (quantified by $\log\Sigma_{M}$) using the circles of different colors. These six environment bins correspond to those used in Section \ref{sec:results}. The units of $\Sigma_{M}$ are $M_{\odot}\,$Mpc$^{-2}$.
\label{fig:cmd}}
\end{figure}

\section{Environments of Galaxies}\label{sec:env}

In this study, we use two definitions to characterize the environments of galaxies. One is the surface number density of galaxies, calculated within the distance to the 10th-nearest galaxy, and another is the surface stellar mass density, determined using the 10 nearest neighboring galaxies. Both are calculated using NSA catalog galaxies with $\log(M_\mathrm{star}/M_{\odot})\ge9.4$ within a rest-frame velocity window of $\pm1000$ km s$^{-1}$ centered on the redshift of each galaxy in the massive ETG sample.\footnote{Therefore, the redshift range of galaxies for computing environments ($0.006<z<0.044$) is slightly broader than that of the parent sample for massive ETGs.}

The surface number density is computed by 
\begin{equation}
\Sigma_{10}=\frac{10}{\pi {r_{10}}^2},
\label{eq:denn}
\end{equation}
in which $r_{10}$ represents the projected distance to the 10th-nearest galaxy, and the units of $r_{10}$ are megaparsecs (Mpc) in this study. Thus, the units of $\Sigma_{10}$ are Mpc$^{-2}$ throughout this study. The surface stellar mass density is then calculated by
\begin{equation}
\Sigma_{M}=\frac{\Sigma_{i=1}^{10} M_{\mathrm{star}, i}}{\pi {r_{10}}^2},
\label{eq:denm}
\end{equation}
in which $M_{\mathrm{star}, i}$ is the stellar mass of the $i$th-nearest galaxy. The units of $\Sigma_{M}$ are $M_{\odot}\,$Mpc$^{-2}$ in this study.

The densities $\Sigma_{10}$ and $\Sigma_{M}$ each have their own advantages and disadvantages. Densities estimated based on the number of galaxies are unaffected by uncertainties in stellar mass measurements, which can show large systematic errors in some cases as mentioned in Section \ref{sec:sample}. However, these number densities do not differentiate between massive galaxies and less massive ones, treating them equally, even though galaxies with different masses might have different impacts on their neighboring galaxies. By contrast, densities based on the stellar mass of galaxies have opposite properties. They take into account the stellar masses of neighboring galaxies but are subject to uncertainties in mass measurements. Thus, we use both density parameters throughout this study to mitigate the possible limitations inherent in each method of defining environments.

We compare $\Sigma_{10}$ and $\Sigma_{M}$ for 40,515 galaxies with $\log(M_\mathrm{star}/M_{\odot})\ge9.4$ within $0.01<z<0.04$ in Figure \ref{fig:den}. A high Spearman's correlation coefficient (0.95) between $\Sigma_{10}$ and $\Sigma_{M}$, coupled with a low scatter (0.17 dex) of data points around the relation between them (see the caption of Figure \ref{fig:den}), demonstrates a strong correlation between the two quantities.\footnote{This suggests that the stellar mass function of galaxies maintains a consistent shape across diverse environments.} For this reason, as demonstrated in Section \ref{sec:results}, the results based on $\Sigma_{10}$ and $\Sigma_{M}$ are almost identical.\footnote{We note that using the surface stellar mass density within an aperture with a radius of 1 Mpc also yields nearly the same results.}

Figure \ref{fig:den} also displays the normalized distributions of $\log\Sigma_{10}$ and $\log\Sigma_{M}$ for both the massive ETG sample and the 40,515 galaxies. These distributions show that massive ETGs with $\log(M_\mathrm{star}/M_{\odot})>11.2$ are more likely to reside in higher-density environments compared to the overall galaxy sample. 

In order to check the basic properties of massive ETGs in different environments, we examine their typical colors and stellar masses. For this purpose, we display the $u-r$ color--stellar mass diagram\footnote{The color values used here are based on $k$-corrected magnitudes derived from two-dimensional S{\'e}rsic fits.} of all galaxies within $0.01<z<0.04$ in Figure \ref{fig:cmd}, on which the sample of massive ETGs with $\log(M_\mathrm{star}/M_{\odot})>11.2$ is also plotted. The figure shows that massive ETGs with $\log(M_\mathrm{star}/M_{\odot})>11.2$ are distributed at the most massive end of the red sequence. The median $u-r$ color of massive ETGs is nearly uniform across different environments, exhibiting a small variation of only 0.03 around 2.59, without any significant trend. In contrast, the typical stellar mass of massive ETGs tends to be slightly higher in denser environments. The median $\log(M_\mathrm{star}/M_{\odot})$ for massive ETGs differs by 0.14 dex between the lowest- and highest-density environments, reaching $\log(M_\mathrm{star}/M_{\odot})=11.41$ in the highest-density environments.

The third rows of Figures \ref{fig:ex_n}--\ref{fig:ex_4} display two-dimensional maps illustrating the spatial distribution of galaxies within $\sim1$ Mpc of each massive ETG. ETGs with tidal features are arranged from Figures \ref{fig:ex_1} to \ref{fig:ex_4}, and ETGs without tidal features are displayed in the panels from left to right in Figure \ref{fig:ex_n}, both in descending order of galaxy density of their environments. The spatial distribution maps in these figures show that the high-density environments with $\Sigma_{10}>48$ (or $\log\Sigma_{M}>11.9$) represent environments like clusters or rich groups, typically containing tens to hundreds of galaxies within a radius of $\sim1$ Mpc. Similarly, the environments with $2.7<\Sigma_{10}<11$ (or $10.7<\log\Sigma_{M}<11.6$) correspond to group-scale environments or regions outside of dense environments, usually hosting a few dozen galaxies within a radius of $\sim1$ Mpc. In contrast, the environments with $0.8<\Sigma_{10}<1.4$ (or $10.2<\log\Sigma_{M}<10.6$) represent poor groups or field environments, characterized by a low galaxy density with only a few galaxies within a radius of $\sim1$ Mpc. Finally, the environments with $\Sigma_{10}<0.4$ (or $\log\Sigma_{M}<10.0$) indicate nearly isolated, low-density environments, in which there are no galaxies or just one galaxy located within a radius of $\sim1$ Mpc.
\\

\section{Detection of Tidal Features}\label{sec:tidal}

A limitation in directly studying galaxy mergers through observational data stems from our inability to observe continuous moments of cosmic time; instead, we only capture a snapshot of the universe. Specifically, this limitation complicates the determination of whether an observed galaxy has recently undergone a merger. However, we can partially overcome this limitation by using tidal features, which are remnants of stellar debris produced by galaxy mergers \citep{Quinn1984,Barnes1988,Hernquist1992,Feldmann2008}. Tidal features are widely used as the most direct observational evidence of mergers that recently occurred. For this reason, many observational studies have been relied on tidal features to identify galaxies that have experienced recent mergers \citep{Schweizer1992,Tal2009,Kaviraj2011,Sheen2012,Duc2015,Hong2015,Sheen2016,YL2020,Yoon2022,Yoon2023a,Bilek2023}. These studies have utilized deep images to detect tidal features, since these features are generally fainter than the main bodies of galaxies.

To detect tidal features, we utilize images from the DESI Legacy Survey Data Release 10 \citep{Dey2019}. The DESI Legacy Survey consists of three wide-area surveys: the Dark Energy Camera Legacy Survey, the Beijing--Arizona Sky Survey, and the Mayall $z$-band Legacy Survey. These three surveys cover an area of $\sim14,000\,\mathrm{deg}^2$ in total. The DESI Legacy Survey images of the $g$ and $r$ bands have a median surface brightness limit of $\sim27$ mag arcsec$^{-2}$, which is determined by $1\sigma$ of the background noise within a $1\arcsec\times1\arcsec$ region. This limit is comparable to that of the deep coadded $r$-band images of the Stripe 82 region in SDSS \citep{YL2020,Yoon2022,Yoon2023a}, which have been frequently utilized for detecting low surface brightness tidal features around galaxies (e.g., \citealt{Kaviraj2010,Schawinski2010,Hong2015}).

Visual inspection is conducted on the $g$- and $r$-band images, as well as on composite color images combining the $g$, $r$, and $z$ bands, to detect tidal features. When scrutinizing the inner regions of galaxies, if needed, we also inspect the residual images of the $g$ and $r$ bands, produced by subtracting the two-dimensional light models of galaxies\footnote{A de Vaucouleurs model, an exponential disk model, and a combination of both models are employed.} from the original images. In the process of visually inspecting the images, we adjust the pixel value scales and enhance signals by applying Gaussian smoothing with various kernel sizes to improve the identification of diffuse and faint tidal features. During this process, we identify 15 massive ETGs located in close proximity to bright stars. As mentioned in Section \ref{sec:sample}, these are excluded from our study owing to the difficulty of discerning tidal features amid the high background illumination caused by these luminous nearby objects.

\begin{figure*}
\includegraphics[width=\linewidth]{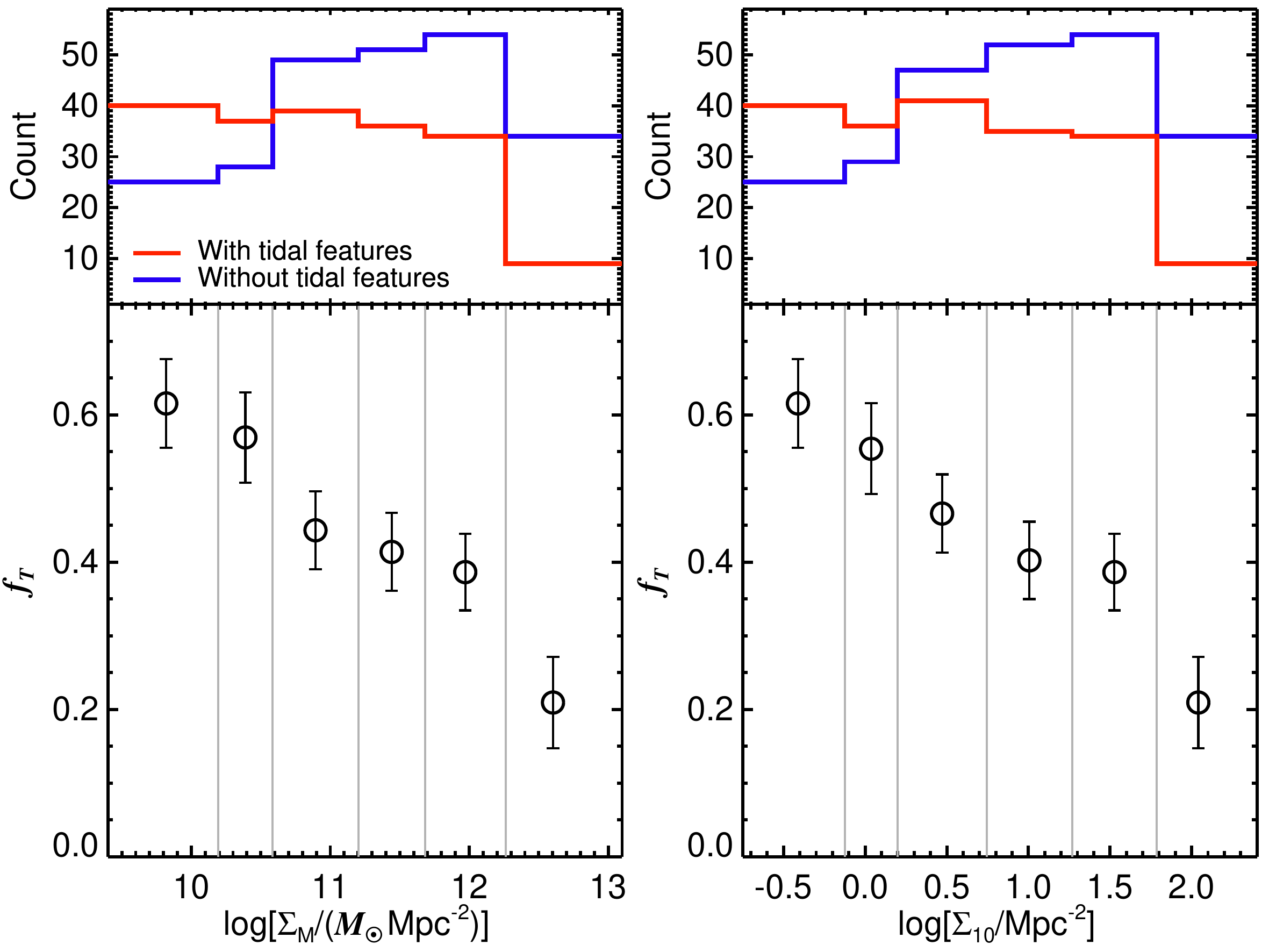}
\centering
\caption{Fraction of massive ETGs with tidal features ($f_T$) as a function of $\log\Sigma_{M}$ (left panel) and $\log\Sigma_{10}$ (right panel). The fraction $f_T$ is calculated in six different environment bins corresponding to the percentile ranges of 0\%--15\%, 15\%--30\%, 30\%--50\%, 50\%--70\%, 70\%--90\%, and 90\%--100\% for both $\Sigma_{M}$ and $\Sigma_{10}$ of massive ETGs with $\log(M_\mathrm{star}/M_{\odot})>11.2$. The gray vertical lines represent the boundaries of the bins. The error bar indicates the standard error of the proportion. Displayed in the top panels are histograms of the distributions of $\log\Sigma_{M}$ and $\log\Sigma_{10}$ for ETGs with tidal features and those without tidal features.
\label{fig:frac_env}}
\end{figure*}

\begin{figure*}
\includegraphics[width=\linewidth]{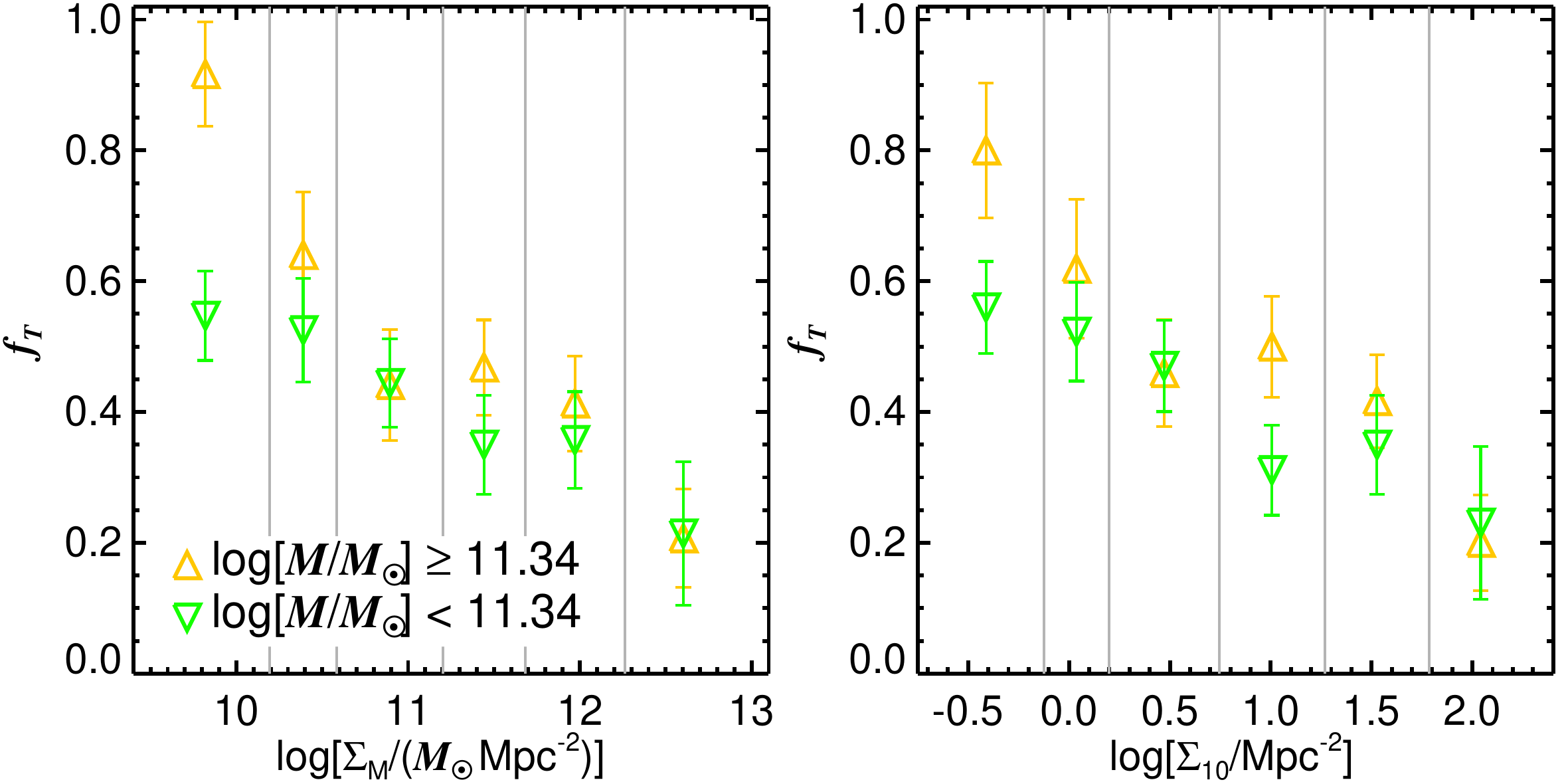}
\includegraphics[width=\linewidth]{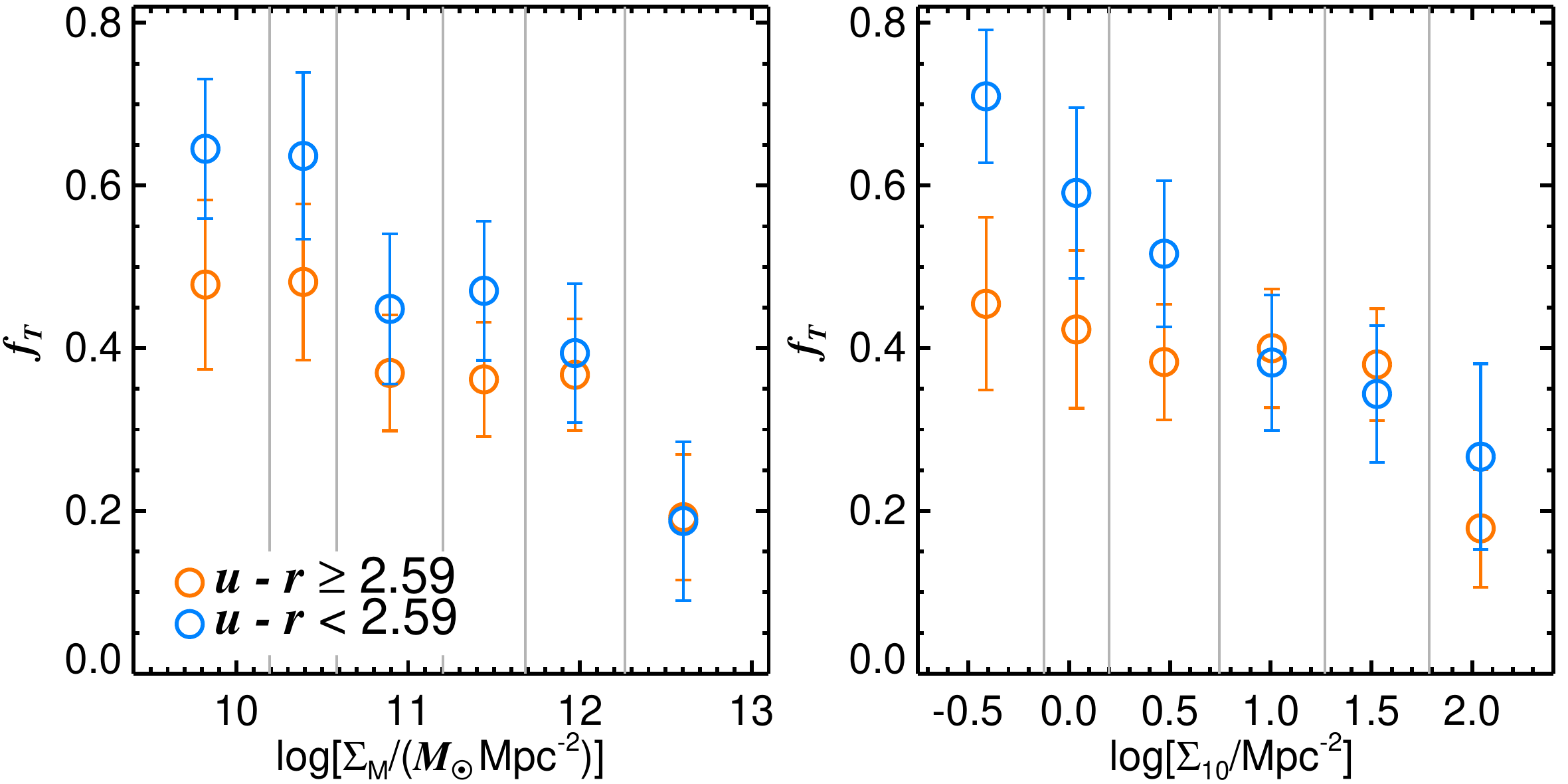}
\centering
\caption{Top panels: fraction of massive ETGs with tidal features ($f_T$) as a function of $\log\Sigma_{M}$ and $\log\Sigma_{10}$ for two categories of massive ETGs, divided by a $\log(M_\mathrm{star}/M_{\odot})$ threshold of 11.34, which approximately corresponds to the median $\log(M_\mathrm{star}/M_{\odot})$ of the massive ETG sample. Bottom panels: $f_T$ as a function of $\log\Sigma_{M}$ and $\log\Sigma_{10}$ for two categories of massive ETGs, separated by a $u-r$ color threshold of 2.59, which is the average $u-r$ color of all the massive ETGs. In the bottom panels, we exclude ETGs with dust lanes that could significantly bias the color values. The fraction $f_T$ is calculated in the six different environment bins, as described in the caption of Figure \ref{fig:frac_env}. The gray vertical lines denote the boundaries of the bins. The error bar represents the standard error of the proportion.
\label{fig:frac_pro}}
\end{figure*} 

Tidal features typically show shapes of tails, streams, or multiple shells \citep{Duc2015,Mancillas2019,Bilek2020,Bilek2023,Sola2022}. Tidal tails are thick, elongated structures that visibly extend from the host galaxies. These elongated stellar features, which may form during major mergers, resemble tidal streams in morphology but are thicker and can sometimes extend to the size of the host galaxy itself. However, in some cases, tails and streams are not obviously distinguishable from each other \citep{Bilek2020,Sola2022}. Tidal streams are narrow, elongated structures that generally resemble thin filaments and are likely associated with minor mergers. In some instances, these stream structures are directly connected to smaller companion galaxies. Tidal shells exhibit arc-shaped structures with sharp edges. Arc features can either align along a common axis or be distributed randomly around the host galaxy. Shell structures at larger distance from the galaxy are usually more diffuse. Some galaxies exhibit more than two types of tidal features. In this study, these tidal tails, streams, and shells around ETGs are defined as tidal features.\footnote{For example, the ETGs in the second and fourth columns of Figure \ref{fig:ex_1} and in the second column of Figure \ref{fig:ex_3} show shell-type tidal features. The ETGs in the first column of Figure \ref{fig:ex_1} and the second column of Figure \ref{fig:ex_4} exhibit tidal streams. The ETGs in the first column of Figure \ref{fig:ex_3} and the fourth column of Figure \ref{fig:ex_4} have tidal tails.} The surface brightness of the tidal features detected in this study is generally brighter than $\sim27$ mag arcsec$^{-2}$, which is the median surface brightness limit of DESI Legacy Survey images.

 We discover that 195 out of 436 massive ETGs with $\log(M_\mathrm{star}/M_{\odot})>11.2$ have tidal features ($44.7\%$). The examples of the deep $r$-band DESI Legacy Survey images for massive ETGs without tidal features are displayed in the second row of Figure \ref{fig:ex_n}, while those for massive ETGs with tidal features are shown in the second rows of Figures \ref{fig:ex_1}--\ref{fig:ex_4}. 

Through the visual inspection, we also classify ETGs that have clear dust lanes, whose potential origin is known to be gas-rich mergers \citep{Oosterloo2002,Kaviraj2012,Shabala2012}. Dust lanes are detected in 53 of 436 massive ETGs ($12.2\%$).\footnote{This fraction may be a lower limit owing to the line-of-sight inclination angles of galaxies.} The ETG in the first column of Figure \ref{fig:ex_2}, as well as the ETGs in the first, third, and fourth columns of Figure \ref{fig:ex_3}, have dust lanes.

In \citet{YL2020} and \citet{Yoon2022}, we have tested our method for identifying tidal features on ETGs in the Stripe 82 region of SDSS by comparing our classifications with those established by \citet{Kaviraj2010}, who categorized ETGs at $z<0.05$ in the Stripe 82 region into normal ETGs and ETGs with tidal features. The test indicates that over $\sim90\%$ of the classifications are consistent with each other (for further details, see \citealt{YL2020} and \citealt{Yoon2022}). The classification of ETGs with dust lanes has also been tested in \citet{YL2020} and \citet{Yoon2022} by comparing our classifications with those of \citet{Kaviraj2010}. This test demonstrates that our identifications of dust lanes are in agreement with those of \citet{Kaviraj2010} after correcting for obvious misclassifications in their work (e.g., all ETGs with dust lanes in our classifications are also classified as such by \citealt{Kaviraj2010} after the correction), though our criteria for dust lane classifications are likely more stringent (refer to \citealt{YL2020} and \citealt{Yoon2022} for more details).

The surface brightness limit may depend on galaxy environments, potentially biasing our results by making it difficult to identify tidal features, for instance, for ETGs in high-density environments. However, we find that the median surface brightness limits for the $g$- and $r$-band images are nearly constant within $\sim0.1$ mag arcsec$^{-2}$ across the six environment bins shown in Sections \ref{sec:env} and \ref{sec:results}. The median image depth for the highest-density environments is shallower by $\sim0.1$ mag arcsec$^{-2}$ compared to that of other environments. However, this difference in depth has negligible impact on the detection of tidal features. This is confirmed by a test in which the depth of images of ETGs with tidal features is artificially reduced by 0.1 mag arcsec$^{-2}$ through the addition of random noise.

For a further test on image depth, we derive our findings as in Section \ref{sec:results}, including only 137 ETGs with prominent tidal features in the category of ETGs with tidal features. The tidal features in these 137 ETGs are clearly visible without requiring smoothed or residual images, or can also be seen in shallow images of SDSS. By doing so, we find a trend nearly identical to the original one presented in Figure \ref{fig:frac_env}.\footnote{Specifically, the fraction $f_T$ is lower in higher-density environments, and in the lowest-density bin, $f_T$ is three times higher than in the highest-density bin.} This implies that the potential discrepancy in detecting tidal features across different environments, due to slight variations in image depth or quality, cannot fundamentally alter our main conclusion.
\\

\section{Results}\label{sec:results}

Our primary result is shown in Figure \ref{fig:frac_env}, which displays the fraction of massive ETGs with tidal features ($f_T$) as a function of $\log\Sigma_{M}$ and $\log\Sigma_{10}$\footnote{Surface stellar mass density and surface number density of galaxies, respectively, which are determined using the 10 nearest galaxies within a rest-frame velocity slice of $\pm1000$ km s$^{-1}$.} calculated across six different environment bins (see the caption). The fraction $f_T$ is defined as $f_T=N_T/N_\mathrm{ETG}$, where $N_T$ is the number of ETGs with tidal features and $N_\mathrm{ETG}$ is the total number of ETGs, regardless of the presence of tidal features. The figure shows a clear trend that $f_T$ is lower in higher-density environments. The fraction $f_T$ in the highest-density bin ($\log\Sigma_{M}\geq12.3$ or $\log\Sigma_{10}\geq1.8$) is $0.21\pm0.06$, while in the lowest-density bin ($\log\Sigma_{M}<10.2$ or $\log\Sigma_{10}<-0.1$) it is $0.62\pm0.06$, which is three times higher than the $f_T$ in the highest-density bin. In the environment bins between the two extremes, $f_T$ gradually decreases as $\Sigma_{M}$ and $\Sigma_{10}$ increases.

This trend is also reflected in the histograms showing the distributions of $\log\Sigma_{M}$ and $\log\Sigma_{10}$ for ETGs with tidal features and those without tidal features, which are displayed in the top panels of Figure \ref{fig:frac_env}. To estimate the statistical significance of the difference in the distributions of $\Sigma_{M}$ and $\Sigma_{10}$ between ETGs with tidal features and those without, we perform Kolmogorov--Smirnov (KS) tests on the distributions of $\Sigma_{M}$ for the two ETG categories, and similarly on the distributions of $\Sigma_{10}$. These tests show that the probability ($0\leq p\leq1$) of the null hypothesis, in which the two distributions stem from the same distribution, is $p=5.5\times10^{-4}$ for $\Sigma_{M}$ and $p=3.8\times10^{-4}$ for $\Sigma_{10}$. This reveals that the environments of massive ETGs with tidal features are significantly different from those of ETGs without tidal features, being biased toward lower densities.

The main trend, that massive ETGs have tidal features more frequently in lower-density environments, is more pronounced for the sample of more massive ETGs with $\log(M_\mathrm{star}/M_{\odot})\ge11.34$ than for ETGs with $11.2<\log(M_\mathrm{star}/M_{\odot})<11.34$, as shown in the top panels of Figure \ref{fig:frac_pro}. The threshold of $\log(M_\mathrm{star}/M_{\odot})=11.34$ approximately corresponds to the median $\log(M_\mathrm{star}/M_{\odot})$. The figure illustrates that both ETG categories display similar $f_T$ values and trends in intermediate- and high-density environments. However, in the lowest-density environments,  $f_T$ is $\approx0.55$ for ETGs with $11.2<\log(M_\mathrm{star}/M_{\odot})<11.34$, whereas a significant majority of ETGs with $\log(M_\mathrm{star}/M_{\odot})\ge11.34$ exhibit tidal features with $f_T=0.80\pm0.10$ or even $0.92\pm0.08$, a fraction that is more than 1.4 times higher than that of the less massive ETGs in the same environments.

Excluding ETGs with dust lanes, which may significantly bias color values, we divide massive ETGs into two categories based on a $u-r$ color threshold of 2.59, which corresponds to the average $u-r$ color for all the massive ETGs (refer to Figure \ref{fig:cmd}). The bottom panels of Figure \ref{fig:frac_pro} exhibit the results. Both red and blue massive ETGs show a similar $f_T$ of $\sim0.2\pm0.1$ in the highest-density environments. By contrast, in the lowest-density environments, blue ETGs exhibit $f_T$ values of $0.65\pm0.09$ and $0.71\pm0.08$, while red ETGs have a value of $\sim0.47\pm0.10$. This suggests that bluer ETGs contribute more significantly to the increased occurrence of tidal features (or the elevated $f_T$) observed in low-density environments. We note that this finding still holds true even for the subcategories of more and less massive ETGs divided by $\log(M_\mathrm{star}/M_{\odot})=11.34$.

The result, showing that the trend of $f_T$ being higher in lower-density environments is more significant for bluer ETGs, is also confirmed by KS tests on the distributions of $\Sigma_{M}$ and $\Sigma_{10}$. Each test is conducted on a pair of the distributions for red ETGs, one with tidal features and one without, and likewise on a pair of the distributions for blue ETGs. By conducting these tests, we find that the KS tests on blue ETGs with $u-r<2.59$ result in $p=1.8\times10^{-2}$ for $\Sigma_{M}$ and $p=3.2\times10^{-3}$ for $\Sigma_{10}$. Meanwhile, the tests on red ETGs with $u-r\ge2.59$ yield $p=0.31$ for $\Sigma_{M}$ and $p=0.21$ for $\Sigma_{10}$. 

As mentioned in Sections \ref{sec:env} and \ref{sec:tidal}, the median $u-r$ color of our ETGs and the image depth are nearly identical across the six environment bins. Moreover, the redshift distributions of our ETGs are also almost the same across the six environments, with the average redshifts differing by less than 0.003 between the environments. Likewise, the median image depth and redshift are nearly identical for ETGs with $\log(M_\mathrm{star}/M_{\odot})\ge11.34$ and ETGs with $11.2<\log(M_\mathrm{star}/M_{\odot})<11.34$, as well as for red ETGs with $u-r\ge2.59$ and blue ETGs with $u-r<2.59$. These facts support the idea that our results shown in Figures \ref{fig:frac_env} and \ref{fig:frac_pro} are not due to differences in factors other than the environments.

\begin{figure*}
\includegraphics[width=\linewidth]{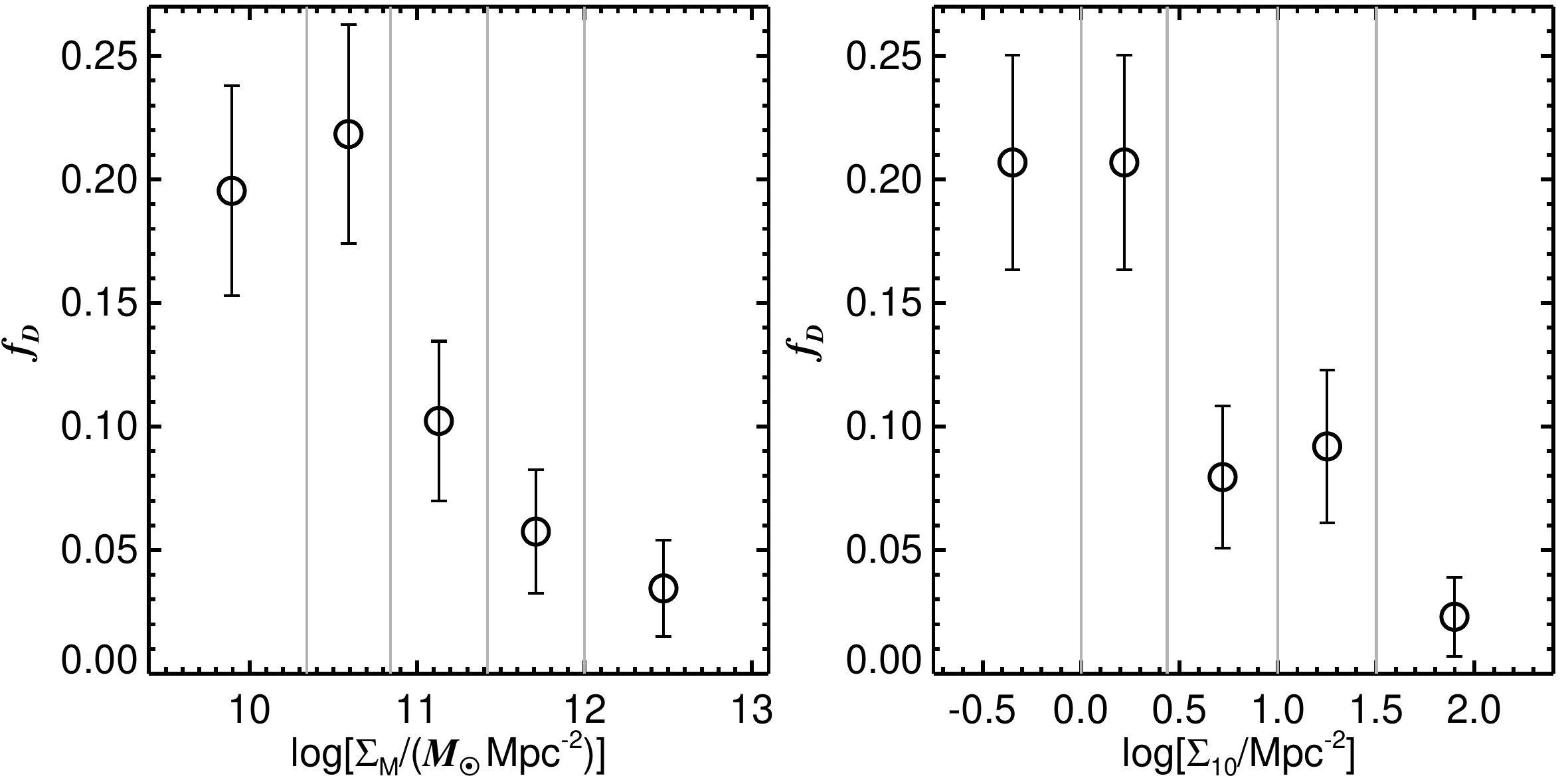}
\centering
\caption{Fraction of massive ETGs with dust lanes ($f_D$) as a function of $\log\Sigma_{M}$ and $\log\Sigma_{10}$. The fraction $f_D$ is computed in five different environment bins corresponding to the percentile ranges of 0\%--20\%, 20\%--40\%, 40\%--60\%, 60\%--80\%, and 80\%--100\% for both $\Sigma_{M}$ and $\Sigma_{10}$ of massive ETGs with $\log(M_\mathrm{star}/M_{\odot})>11.2$. The gray vertical lines denote the boundaries of the bins. The error bar represents the standard error of the proportion.
\label{fig:frac_dust}}
\end{figure*}

The fraction of ETGs with dust lanes ($f_D$) as a function of $\log\Sigma_{M}$ and $\log\Sigma_{10}$ is also examined, and the results are displayed in Figure \ref{fig:frac_dust}. The figure shows a clear trend of $f_D$ being higher in lower-density environments. In the highest-density bins with $\log\Sigma_{M}\geq12.0$ and $\log\Sigma_{10}\geq1.5$, the $f_D$ values are $0.03\pm0.02$ and $0.02\pm0.02$, respectively. By contrast, in the low-density environments with $\log\Sigma_{M}<10.8$ or $\Sigma_{10}<0.4$, $f_D$ is $\sim0.21\pm0.04$, which is more than six times higher than $f_D$ in the highest-density bin. By performing KS tests on the distributions of $\Sigma_{M}$ for ETGs with and without dust lanes, and similarly on the distributions of $\Sigma_{10}$, we obtain $p=1.1\times10^{-4}$ for $\Sigma_{M}$ and $p=1.0\times10^{-5}$ for $\Sigma_{10}$. This indicates that the environments of massive ETGs with dust lanes differ significantly from those of ETGs without dust lanes, showing a bias toward lower densities.

We find that $66.0\%$ of ETGs with dust lanes have tidal features, a fraction that is 1.6 times higher than that of ETGs without dust lanes ($41.8\%$). ETGs with dust lanes in the lowest-density bin of Figure \ref{fig:frac_dust} exhibit $f_T=0.76$, and high values of $f_T\gtrsim0.6$ are observed in ETGs with dust lanes across all environments. The high $f_T$ values detected in ETGs with dust lanes indicate that dust lanes in ETGs are associated with recent merger events, as suggested by \citet{YL2020}.

In conclusion, our results demonstrate that the most massive ETGs with $\log(M_\mathrm{star}/M_{\odot})>11.2$ exhibit tidal features more frequently in lower-density environments. This trend is more substantial for more extremely massive ETGs and is predominantly driven by ETGs having bluer $u-r$ colors. In addition, these massive ETGs are more likely to have dust lanes in lower-density environments.
\\

\section{Discussion}\label{sec:discuss}

One explanation for our main finding---that the most massive ETGs have tidal features more commonly in low-density environments---is that such ETGs in lower-density environments have genuinely experienced recent mergers (which occurred $\sim2$--$4$ Gyrs ago; \citealt{Ji2014,Mancillas2019,YL2020}) more frequently than their counterparts in higher-density environments. This fits in with the general expectations of the current cosmological model for galaxy formation \citep{Kauffmann1996,Gottlober2001,DeLucia2006,Niemi2010,Yoon2017,Yoon2023b}, which are outlined in Section \ref{sec:intro}, suggesting that these expectations are also valid for ETGs at the most massive end. 

Within the framework of this model, our results are interpreted to mean that a large portion of the most massive ETGs in high-density environments, such as galaxy clusters at low redshifts, have not undergone significant recent mergers that leave tidal features, due to the high relative velocities among galaxies in fully developed, massive halos.\footnote{Instead, massive ETGs in high-density environments are more likely to have experienced frequent mergers and consequent evolution already at an early epoch of $z\gtrsim2$, when galaxies were moving with low velocities in immature halos, resulting in an increased merger rate.} Some of the tidal features detected in cluster environments may originate from mergers that occurred in prior, smaller halos or in the outskirts of clusters before entering the current clusters \citep{Yi2013,Kim2024}.

 In contrast, within the framework of the model, our results also suggest that massive ETGs in low-density or even isolated environments have prolonged assembly histories extending to the current epoch. This is most evident in ETGs at the extremely massive end, as over $80\%$ of those in the lowest-density environments exhibit tidal features. This may align with previous studies arguing that a large and luminous ETG in an isolated environment can originate from the collapse or merger of a compact, dense galaxy group \citep{Barnes1989,Mulchaey1999,Reda2004}.

Considering that bluer massive ETGs are more responsible for the elevated $f_T$ in low-density environments, and that dust lanes are more commonly observed in massive ETGs in these environments, it is likely that the increased rate of recent mergers in low-density environments, compared to the rate in high-density environments, is primarily attributable to gas-abundant mergers. This is because mergers with abundant gas can effectively result in a bluer $u-r$ color by promoting the formation of young stars \citep{Hernquist1989,Mihos1996,Springel2005}, with dust lanes likely being remnants of such star-forming activity \citep{Oosterloo2002,Kaviraj2012,Shabala2012}.

Another factor that can explain our main finding is that tidal features may be more easily destroyed or have shorter lifetimes in denser environments, as noted in previous studies \citep{Colbert2001,Kuntschner2002,Adams2012,Ji2014,Rampazzo2020}. This can be due to continuous gravitational interactions with a cluster potential and nearby neighbors passing around a massive ETG, or because stars that make up tidal features easily exceed the escape velocity during merger processes involving high-velocity encounters in dense environments. 

For instance, \citet{Adams2012} demonstrated that the observed scarcity of tidal features at the centers of clusters can be partially explained by the reduced lifetimes of these features in denser environments, which result from shorter dynamical times. \citet{Ji2014} suggested that the duration time of tidal features is shortened by $30\%$ owing to the tidal forces of a cluster potential, which can effectively strip away such features.

To determine which of the two explanations---(1) merger histories and (2) the lifespan of tidal features---is more responsible for our results and to assess the extent of their respective contributions, we need further intensive examinations based on simulations and deeper images from large surveys. For example, in cosmological simulations, we can identify very massive ETGs, both with and without tidal features, located in environments ranging from isolated to extremely dense, and trace their respective merger histories. By doing so, we can verify our results and their origins. Although \citet{Khalid2024} conducted a study with a partially similar concept,\footnote{The result of \citet{Khalid2024} aligns with ours in that tidal features are less abundant in galaxies within more massive halos, when halo masses are larger than $\sim10^{12}\,M_{\odot}$.} it is more preferable to use cosmological simulations with larger volumes and higher resolutions, in order to accurately identify tidal features, even such as small and faint ones, and to secure the statistically sufficient number of the most massive ETGs in various environments, including extreme ones.

Future large survey images with a surface brightness limit deeper than $\sim29$ mag arcsec$^{-2}$ will allow us to identify extremely faint tidal features in many galaxies across diverse environments. These very faint features can observationally trace mergers that occurred a long time ago ($\gtrsim4$ Gyr).\footnote{For instance, \citet{vanDokkum2005} showed that a large fraction ($71\%$) of nearby ETGs exhibit signatures of tidally disturbed features, using very deep images with a surface brightness limit of $\sim29$ mag arcsec$^{-2}$.} Thus, such deep image data can be used to support the analysis of the merger histories of massive ETGs in various environments that are obtained from cosmological simulations.

Finally, the lifetime of tidal features should be intensively tested using $N$-body or hydrodynamic simulations, by examining tidal features originating from an equivalent set of mergers in simulated clusters, groups, and an isolated environment. For example, \citet{Ji2014} conducted simulations and observed shortened lifetimes of tidal features within a cluster potential. However, it is crucial to understand not only the effect of a cluster potential alone but also the impact of multiple interactions with nearby galaxies on tidal features in densely populated cluster and group environments.
\\

\section{Summary}\label{sec:summary}

Using the 436 most massive ETGs with $\log(M_\mathrm{star}/M_{\odot})>11.2$ at $0.01<z<0.04$, we examine the variation in the fraction of massive ETGs with tidal features ($f_T$) across different environments and verify whether the most massive galaxies commonly display tidal features in very low density environments. The environments are defined as the surface number density or surface stellar mass density of galaxies, determined using the 10 nearest galaxies within a rest-frame velocity slice of $\pm1000$ km s$^{-1}$. We detect tidal features, which serve as direct evidence of recent mergers, through a visual inspection of DESI Legacy Survey images whose depth is sufficient for the identification of such features.

Our main finding is that the most massive ETGs exhibit tidal features more frequently in lower-density environments. The fraction $f_T$ is $0.21\pm0.06$ in the highest-density environments, where tens to hundreds of galaxies are within a radius of $\sim1$ Mpc. By contrast, in the lowest-density environments with only a handful of neighboring galaxies or fewer, $f_T$ is $0.62\pm0.06$, which is three times higher than $f_T$ in the highest-density environments. This trend is more prominent for more extremely massive ETGs with $\log(M_\mathrm{star}/M_{\odot})\ge11.34$, with $f_T$ reaching up to $0.92\pm0.08$ in the lowest-density environments.

One explanation for our finding, within the framework of the current cosmological model for galaxy formation, is that the most massive ETGs in lower-density environments have indeed undergone recent mergers more frequently than their counterparts in higher-density environments, suggesting that they possess more extended formation histories that continue into the present. Another possible explanation for this finding is that tidal features may have shorter lifetimes in denser environments owing to external factors inherent in these environments. 

We also find that massive ETGs with bluer $u-r$ colors are more responsible for the elevated $f_T$ in low-density environments. In addition, the fraction of massive ETGs with dust lanes is more than six times higher in low-density environments than in high-density environments. These imply that the increased rate of recent mergers in low-density environments, compared with the rate in high-density environments, may primarily stem from gas-abundant mergers, provided that the high $f_T$ in low-density environments truly results from a higher rate of recent mergers.
\\

\begin{acknowledgments}
This research was supported by the Korea Astronomy and Space Science Institute under the R\&D program (project No. 2024-1-831-00), supervised by the Ministry of Science and ICT.
\end{acknowledgments}



\end{document}